**Title:** The explanatory power of activity flow models of brain function

**Author:** Michael W. Cole
The Center for Molecular and Behavioral Neuroscience, Rutgers University, Newark, NJ 07102

*A chapter in the book Computational and Network Modeling of Neuroimaging Data (editor: Kendrick Kay)*

# Chapter Summary

Tremendous neuroscientific progress has recently been made by mapping brain connectivity, complementing extensive knowledge of task-evoked brain activation patterns. However, despite evidence that they are related, these connectivity and activity lines of research have mostly progressed separately. Here I review the notable productivity and future promise of combining connectivity and task-evoked activity estimates into activity flow models. These data-driven computational models simulate the generation of task-evoked activations (including those linked to behavior), producing empirically-supported explanations of the origin of neurocognitive functions based on the flow of task-evoked activity over empirical brain connections. Critically, by incorporating causal principles and extensive empirical constraints from brain data, this approach can provide more mechanistic accounts of neurocognitive phenomena than purely predictive (as opposed to explanatory) models or models optimized primarily for task performance (e.g., standard artificial neural networks). The variety of activity-flow-based explanations reported so far are covered here along with important methodological and theoretical considerations when discovering new activity-flow-based explanations. Together, these considerations illustrate the promise of activity flow modeling for the future of neuroscience and ultimately for the development of novel clinical treatments (e.g., using brain stimulation) for brain disorders.

# Glossary

**Activity flow** – The movement of activity between neural populations.
**Activity flow models** – Empirically-constrained simulations of the generation of neurocognitive functions via activity flow processes.
**Functional/effective connectivity (FC)** – FC methods estimate the statistical association between neural time series. The typical theoretical target of FC methods is the causal relationship between the neural populations contributing those time series. However, FC methods vary substantially in their degree of causal validity, with no current FC method thought to yield perfect causal inferences.
**Generative understanding** – A way of comprehending a phenomenon and the system that generates it such that the necessary and sufficient conditions of the system are known for what generates that phenomenon. This is especially useful for predicting the impact of novel configurations of the system on the phenomenon of interest, such as a system break down (e.g., brain disorder) or intervention (e.g., brain treatment or brain enhancement).
**Held-out activity** – Brain activity that is independent from the activity used to build a model, such that a prediction of that activity from the model would be logically non-circular. In the context of activity flow modeling, this refers to independence between the target activity that is to be generated/predicted by a model and the data used to build that model.
**Neurocognitive function** – A neural phenomenon associated with a cognitive or behavioral process, which can be a target of an explanation. Note that function does not imply that the to-be-explained phenomenon is the ultimate purpose of the involved neural populations. Examples: face selectivity in the fusiform face area, behavior being driven by primary motor cortex activity.

**Prediction (versus causality)** – The generation of data that can be compared to a phenomenon of interest, with a better prediction resulting from a better match between the generated and actual phenomenon. Some non-causal (e.g., statistical) models can produce accurate predictions, such that the true causal model/process is only one of a variety of good predictive models. However, most possible (e.g., random) models are not good predictors, making good prediction a necessary but not sufficient condition for a model being the true causal model.
**Resting-state functional connectivity** – FC estimated using functional brain data (such as fMRI) while a participant rests.
**Task-evoked activity** – Activity elicited by experimenter-presented stimuli, such as sensory input and task instructions.

## Introduction to activity flow modeling

Activity flow is defined as the movement of activity between neural populations (Cole et al., 2016). The concept of brain activity flow is everywhere and nowhere in the neuroscience literature. It is everywhere in the sense that the standard model of neural transmission – wherein action potentials flow along axons to influence downstream neurons via neurotransmitter release impacting their dendrites – involves the flow of "activity" (electrochemical signals). Yet activity flow is nowhere in neuroscience in the sense that neuroscientific inferences are normally made using *either* activity patterns or connectivity. Seven years of studies focusing on combining activity patterns and connectivity (typically using functional/effective connectivity (FC)) to build activity flow models – simulations of the generation of neurocognitive functions via activity flow processes – has demonstrated the broad utility of this approach above and beyond standard activity *or* connectivity approaches alone. Modeling integration between task-evoked activity and connectivity to make strong inferences about brain function will likely be essential for developing rich causal explanations of the neural basis of cognitive functions, for the purpose of fundamental understanding and developing treatments for brain disorders.

To better illustrate the relevance of the activity flow modeling approach, let us consider a hypothetical scenario wherein an alien technology lands on Earth: the optimal brain imager (OBI). After some fiddling, human scientists discover that the OBI noninvasively reads out every aspect of the entire human brain at the atomic level (including electromagnetic fields) at microsecond resolution. Immediately, full brain scans of human brain anatomy and human brains performing all manner of tasks are collected, and the data are rapidly analyzed by the accompanying alien computer (capable of handling the massive datasets produced by the OBI). These analyses map the human connectome at full molecular resolution, along with cellular-level whole brain maps of neural activity patterns accompanying every stimulus and task variable. Neuroscientists predict that all of the mysteries of the brain will soon be solved. Certainly, some mysteries are rapidly solved, and neuroscientists rejoice.

However, it eventually becomes clear that the data and analyses derived from the OBI present a major barrier to fundamental understanding of how brain activity generates cognition and behavior. Specifically, even after full mapping of brain connectivity and function-related brain activity, it remains unclear how these "parts" work together to generate neurocognitive functions. Such *generative* understanding of function is akin to understanding how the parts of a car work together (causally interact) to generate the emergent properties of rapid and controlled movement (Ito et al., 2020). For example, understanding how pressing a car's accelerator generates movement requires knowledge of the causal relationships between the system's components (e.g., the pedal, the transmission, the engine, and the wheels). Without such an understanding, cars would appear to be mysterious to us, with the practical problem that no one would be able to fix a car if it were to break down. This is largely the situation with the human brain – many of its functions remain mysterious to us, and we have only a very limited ability to fix it when brain disorders arise.

Activity flow modeling is one possible solution to the problem of generative understanding. Rather than treating mapping of brain connectivity and brain activity patterns as goals in and of themselves, these become starting points in the quest for generative understanding of brain function. Indeed, activity flow models are directly built on the combination of brain connectivity and brain activity patterns: First, empirical connectivity between all nodes (neural populations) of interest are identified, followed by input of empirical task-evoked activity into those nodes. A subset of nodes are "held out" – meaning their empirical task-evoked activities are not used as input – such that the activity of those nodes can instead be generated by the model. The basic generative process is the same as most neural network simulations: a node's activity is determined by the connectivity-weighted sum of the activity inputs into that node (**Figure 1A**). Model accuracy is assessed by comparing the generated task-evoked activity with the empirical task-evoked activity (**Figure 1B**). To the extent that the generated task-evoked activity is accurate for a given neural population, the contributing activity, connectivity, and activity flow (i.e., activity-connectivity interaction) processes can be analyzed to infer details of the generative processes driving task-evoked activity in that neural population.

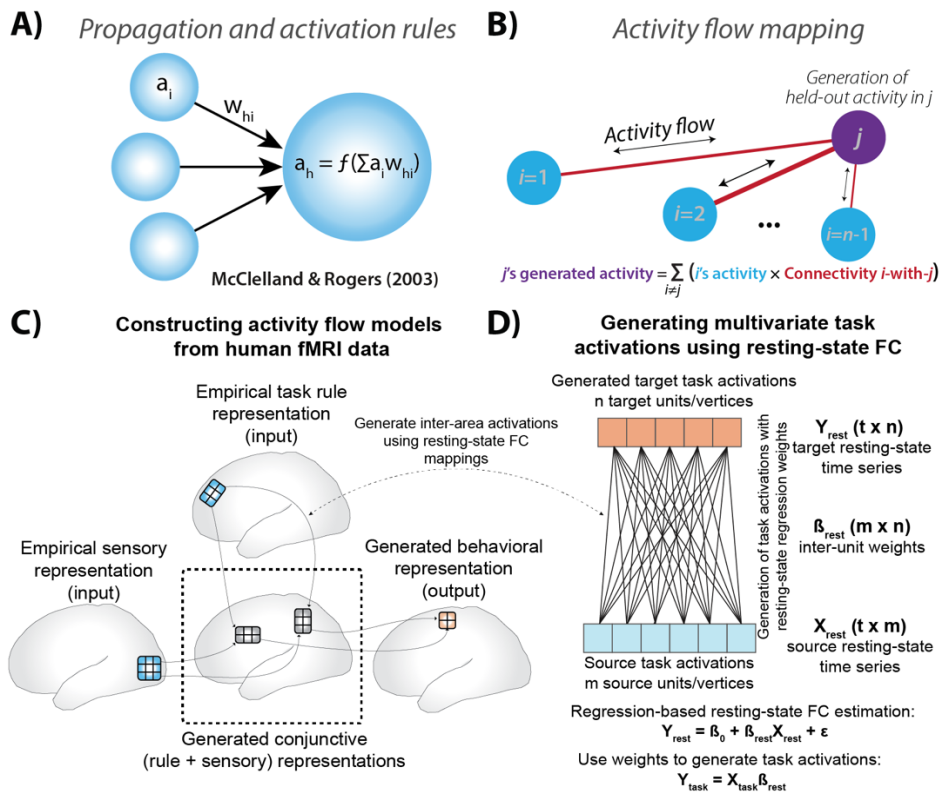

**Figure 1 – Overview of activity flow modeling in terms of theoretical origin, methodology, and practical steps for model building.**
**A**) Theoretical origin of activity flow modeling. The same propagation and activation rules used in connectionist artificial neural networks are used in activity flow models, but the parameters are set from empirical data. Adapted from McClelland & Rogers (2003). **B**) Basic methodological schematic for activity flow modeling. The standard activity flow mapping approach (Cole et al., 2016) involves generating task-evoked activity in each region one-at-a-time based on activity in all other regions and the connectivity with the to-be-generated region. The connectivity with the target region, along with the equation (transfer function) at the bottom of the panel, constitutes that target region's activity flow model. **C**) An example activity flow model based on Ito et al. (2022), going from sensory inputs and task rule representations to motor outputs in a context-dependent decision-making task. This example goes beyond the one-at-a-time activity flow mapping approach in panel B, linking sensory inputs to cognitive transformations and behavior (motor responses). **D**) How to build an activity flow model. Most activity flow models use standard multiple regression to estimate functional/effective connectivity (FC) (model weights) between brain regions. Resting-state brain data is typically used to estimate functional connections, though task data can also be used. These FC patterns are then used (along with the previous layers' task-evoked activity) to generate task-evoked activity in the next layer in the multi-step activity flow procedure. Figure adapted from (Ito et al., 2020) and (Ito et al., 2022).

Going back to the hypothetical OBI, systematic application of activity flow modeling to test generative hypotheses could be an effective strategy for achieving a generative understanding of brain function. For example, generative understanding of how an individual is able to read aloud could involve OBI-based recording of their brain activity during vocalized reading of a long passage of text. This individual's detailed connectome would be used in a network simulation, with activity recorded from the individual's primary auditory cortex during verbal instructions (to read the text) input into the model. The model would then simulate the resulting activity flows moving throughout the brain as the individual prepares for the reading task. The text stimuli would then be input into the simulation via visual activity input (e.g., empirical activity patterns from primary visual cortex during stimulus presentation), with the resulting activity flows throughout the brain mixing with the activity previously initiated during instructions. To the extent that activity patterns following the initial inputs are accurately generated – especially the motor activity patterns directly driving the behavior of interest (reading aloud) – the various simulated activity flows driving those processes can be analyzed for insights into the generation of reading aloud behavior. Further, changes to those activity flows can be made within the model to better understand their impact, resulting in yet deeper understanding of the generative processes involved. Ultimately, activity flow modeling would go beyond the comprehensive neural activity and connectivity information obtained by the OBI to provide insights into how activity and connectivity interact to produce cognition and behavior.

Core principles of activity flow modeling

There are four principles that are central to activity flow modeling (**Table 1**). The first of these principles was described in the previous section: *generativity*, wherein functions of interest (e.g., face selectivity in the fusiform face area, or behavioral responses via primary motor cortex) are produced in the act of modeling brain function. Generating the function of interest allows quantification of the success of the proposed mechanism for producing that function (Ito et al., 2020). Thus, generativity is key to the utility of activity flow modeling in providing evidence regarding the likely brain processes underlying functions of interest. This contrasts with most data analysis approaches (such as estimating connectivity or classifying task-evoked activation patterns), which are descriptive rather than generative.

**Table 1 – Core principles underlying activity flow modeling**

| Activity flow modeling principle | Description |
|---|---|
| Generative | Use activity and connectivity to generate held-out (independent) activity in one or more neural population, allowing assessment of the causal sufficiency of the used activity and connectivity features in generating functions of interest |
| Simplicity/abstraction | Increase interpretability and identification of essential model features through model simplicity and abstraction (e.g., focusing on activity of entire brain regions rather than within-region activity |

|  | patterns), increasing complexity only as necessary based on empirical evidence and the to-be-generated function(s) for a given study |
| --- | --- |
| Mechanistic/causal | Add accurate causal constraints (e.g., to connectivity estimates) whenever possible to increase the likelihood that modeled processes match the causal mechanisms used in the brain |
| Empirically constrained (data driven) | Model features should be directly estimated by empirical brain data when possible, grounding the model in reality and reducing the number of modeling assumptions |

Generativity follows from the central equation in activity flow modeling, which is also used in most artificial neural networks: $a_h = f(\Sigma a_i w_{hi})$, where $a_h$ is the to-be-generated neural population's activity, $w_{hi}$ is the connectivity to region $a_h$, and $f$ is a transfer function such as a sigmoid, linear, or rectified linear (threshold) transformation (**Figure 1A**). Thus, activity flow modeling can be considered an approach to generate empirically-estimated neural network simulations, aiding interpretation of neural data by testing its ability to generate neural data in independent/held-out neural populations (Ito et al., 2020).

Another core principle in the development of activity flow models has been *simplicity/abstraction* – keeping the models as simple as possible until there is an explicit need to make them more complex (**Table 1**). This includes the concept of abstraction, wherein features at lower levels of organization are aggregated over in the interest of explanatory simplicity. For example, we abstract over the quantum and atomic scale when describing the behavior of individual neurons, and so it may be that the most effective generative explanations of neurocognitive functions are at yet larger scales (Saxena & Cunningham, 2019). While simplicity and abstraction typically result in activity flow models lacking many details included in more biophysically detailed models, this simplicity principle has several important advantages. First, simplicity reduces the chance of the models containing a large number of unjustified assumptions that turn out to be incorrect. Instead, we start out simple with new assumptions treated as hypotheses to be supported by data and/or simulations. Second, simplicity can reduce overfitting of model features to particular problems, tasks, or data types, increasing the generality of findings (Hansen, 2020; Li & Spratling, 2023). Third, simplicity is obtained in part by using the methods that most researchers are actively using (e.g., using general linear models (GLMs) for fMRI activity estimates and Pearson correlation for fMRI connectivity estimates), making activity flow modeling results easier for most researchers to understand, as well as maximally relating results to previous studies. Thus, simplicity has been advantageous for activity flow models, even as they become more complex (and thus able to account for more phenomena) with each study.

To briefly illustrate this simple-to-complex transition, let us start with the original activity flow modeling study (Cole et al., 2016), which used standard Pearson-correlation FC and GLMs to maximally relate to the existing fMRI literature. That same study then demonstrated (in both simulations and empirical fMRI data) the increased causal validity and generative performance of using multiple regression as an FC measure. Thus, results were maximally applicable to the existing literature, while advancing the literature using a more complex and causally principled FC approach (multiple-regression FC). This simple-to-complex trajectory has led most recently to a multi-step activity flow approach (Ito et al., 2022) (**Figure 1C & 1D**), using multiple-regression FC across "layers" (sets of brain regions) from 1) visual and auditory inputs to 2) intermediate "conjunction regions" implementing cognitive information transformations to 3) motor outputs in primary motor cortex. These model-generated motor activations are then decoded (based on a decoding model trained using empirical motor activations), such that the model generates task-performing behavior. I will be unpacking the details of these models below, but I briefly described them here to illustrate the scale of the simple-to-complex transition so far (across 17 studies published between 2016 and 2023).

Another core principle of activity flow modeling is to focus on *mechanistic* explanations, by way of identifying causal constraints whenever possible (**Table 1**). The importance of this principle derives from the observation that an explanation provides less generative understanding if the generating/predicting process is unrelated to causal mechanisms within the system of interest. For example, pure (non-mechanistic) prediction of a motor response from a statistical model (e.g., based on the previous history of motor responses) rather than one grounded in causality would provide less insight into the processes generating that motor response than alternate predictions based on causally grounded generative processes.

The mechanistic principle is perhaps the most challenging principle to enact, given the difficulty of causal inference in general and in the context of brain interactions in particular (R. D. Mill, Bagic, et al., 2017; R. D. Mill, Ito, et al., 2017; Reid et al., 2019). The challenges of causal inferences are numerous and complex (see Pearl, 2009), but an example is confounding. To illustrate, consider that pain killer use (such as acetaminophen) is strongly correlated with mortality (Lipworth et al., 2003). This may lead one to avoid pain killer use, yet this is a non-causal relationship due to confounding. Specifically, many diseases (e.g., cancer) cause pain (and so pain killer use) as well as mortality. This scenario can be represented as a simple causal graph (A←C→B), with C being disease, A being pain killer use, and B being mortality. Critically, a causal inference method (e.g., simple correlation) that does not take into account confounding would suggest the wrong conclusion that pain killer use causes death. The same basic problem arises frequently in neuroscience (Reid et al., 2019).

Despite these challenges, causal/mechanistic explanation is a core principle as causality is central to scientific understanding and application of that understanding, such as causal interventions (i.e., treatments) to cure brain diseases. As an example of this principle in action, consider the use of multiple-regression FC in the original activity flow modeling study (Cole et al., 2016), which was based on the improved causal inferences possible using multiple-regression FC (due to accounting for causal confounds and causal chains) relative to the more standard Pearson-correlation FC approach (**Figure 2A & 2B**). Further subsequent progress was made, however, with a recent study demonstrating that multiple-regression FC (and partial-correlation FC) is less accurate in capturing ground truth FC than Pearson-correlation FC in the case of causal colliders (**Figure 2C**) (Sanchez-Romero & Cole, 2021). That same study demonstrated a method – termed combinedFC – that combines the advantages of both multiple-regression FC (or partial-correlation FC) and Pearson-correlation FC, increasing the causal validity of FC beyond either method alone. Further, that study used empirical fMRI data to find massive reductions in the number of estimated connections with partial-correlation FC relative to regular Pearson-correlation FC, suggesting confounders and chains (affecting correlation FC) are a much bigger problem in practice than colliders (affecting multiple regression/partial correlation). Accordingly, if a method like combinedFC (or the Peter-Clark algorithm (Sanchez-Romero et al., 2023)) is not used then multiple-regression FC (or partial-correlation FC) is preferrable to Pearson-correlation FC.

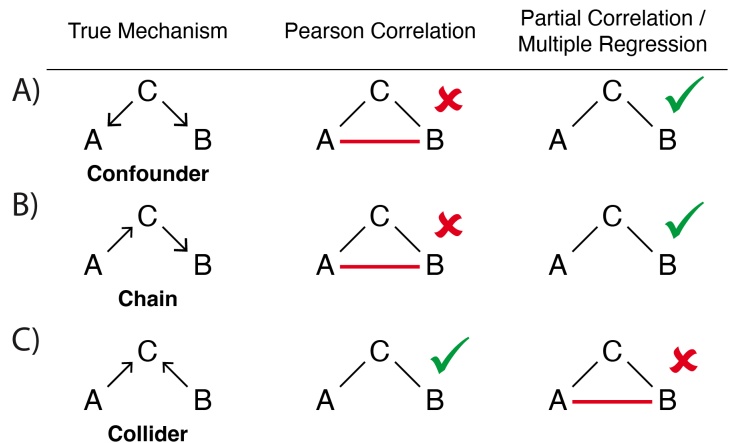

**Figure 2 – The relationship between underlying causal mechanisms and common FC measures.** Three common causal graph motifs are shown to illustrate fundamental principles of relating causal mechanisms to FC measures. CombinedFC (Sanchez-Romero & Cole, 2021) is an FC method that combines the best of both bivariate Pearson correlation and partial-correlation/multiple-regression FC. Other methods that achieve similar (and sometimes better) results also exist (Sanchez-Romero et al., 2023). **A**) A causal confounder (also termed a *fork*) is when a neural population causes activity in two or more others, causing correlations in the downstream populations' time series. This can lead to spurious associations/causal inferences. Common FC measures such as Pearson correlation and coherence are susceptible to false positives from confounders (Reid et al., 2019). **B**) A causal chain is when a neural population influences another via a third neural population. Not accounting for the third neural population (i.e., a mediator) can lead to false positive connections (though these could be considered "indirect" connections). **C**) A causal collider occurs when two or more neural populations influence another neural population. In this case common methods such as Pearson correlation or coherence make proper inferences, but methods that control for confounding (e.g., partial correlation) create false connections. Note that, at least with fMRI data, it was found that the confounding problem (panel A) was much more problematic in the human brain than the collider problem, though both are present (Sanchez-Romero & Cole, 2021). Figure adapted from (Sanchez-Romero & Cole, 2021).

Also building on the mechanistic/causal principle, a recent expansion of activity flow modeling to source-localized high-density electroencephalography (EEG) datasets (R. D. Mill et al., 2022) improves causal inferences of activity flow modeling further by taking into account the temporal order of causal events (**Figure 3**). These approaches have also been combined with simulated lesions and other innovations to improve causal inferences once activity flow models are built (Hearne et al., 2021; Ito et al., 2022; R. D. Mill et al., 2022). Despite these advances in FC methodology and the consequent improvement in mechanistic inferences with activity flow modeling, numerous opportunities to increase the causal validity of activity flow models remain.

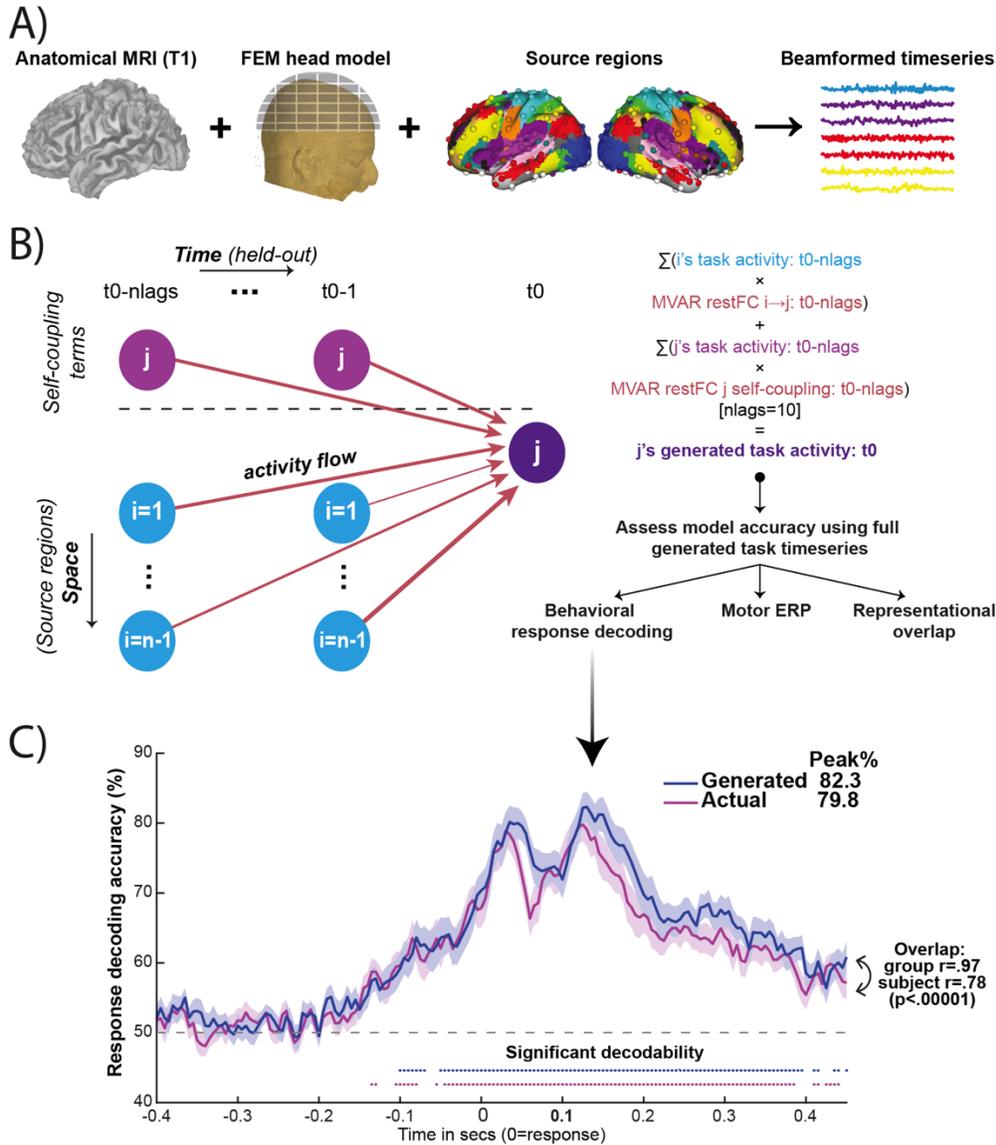

**Figure 3 – Improving causal mechanistic inferences via extension of activity flow modeling/mapping to high-temporal-resolution neural recordings (here EEG).**
Approach and results from Mill et al. (2022). **A**) High-density EEG data in human participants were used along with individual-subject structural MRI to implement region-level source localization. **B**) The dynamic activity flow mapping framework. Instead of just generating activity in spatially held-out regions, this approach also generates *temporally* held-out (i.e., future) activity. This improves the causal inference given the centrality of the direction of time in causal interactions. **C**) The decoded motor response information time course based on generated cortical motor network activity (blue) from an activity flow model (including all non-motor regions) during a simple sensory-motor mapping task. The decoded motor response information time course from the actual cortical motor network activity (pink) is shown for comparison. Figure adapted from (R. D. Mill et al., 2022).

Finally, perhaps the most fundamental principle in activity flow modeling is the centrality of *empirical constraints* (**Table 1**). Indeed, activity flow modeling begins and ends with empirical data, such that it can be considered both a data analysis and computational modeling framework. The approach begins with empirical data in the sense that model features – connections and input activity – are directly estimated from empirical brain data. The approach ends with empirical data in the sense that the model-generated activity is compared to actual empirical brain activity to test the validity of the model for generating the brain function(s) of interest. For example, a recent study (Cocuzza et al., 2022) used a causal FC approach (combinedFC; see **Figure 2**) with fMRI to estimate the empirical connectivity between brain regions in the visual system, then used those empirical connections along with empirical activations in V1 as input into the model. This model was then tested for its ability to generate well-known category-selectivity activity in higher-level visual regions, such as face selectivity in the fusiform face area (**Figure 4**). The success of the model in generating such category-selective activity demonstrated the sufficiency of fMRI connectivity and activity – when combined in an activity flow model – for providing an (important but partial) explanation of how category-selective visual brain activity is generated via brain network interactions.

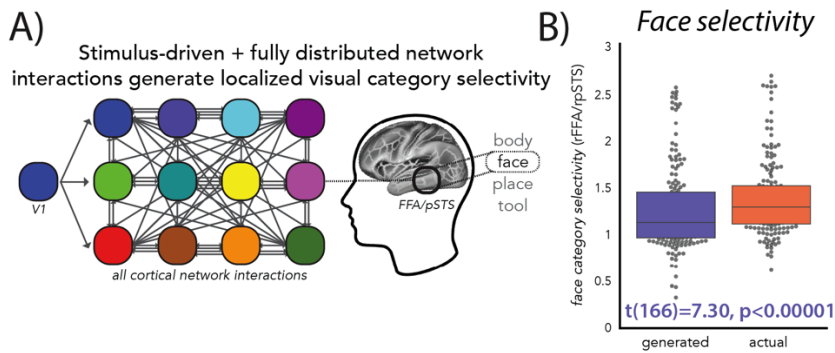

**Figure 4 – Activity flow modeling of visual processing generates canonical visual category selectivity via distributed processes.**
Approach and results from Cocuzza et al. (2022). **A**) A multi-step activity flow model was used, with the only empirical task-evoked activations coming from V1 activity patterns. All other task-evoked activations were generated via activity flow over resting-state FC (with combinedFC). Three activity flow steps were simulated to generate task-evoked activity in visual category selective region sets (panel B). **B**) Generated and actual category selectivity for an example visual category selective region set (fusiform face complex: fusiform face area + other face-selective regions) are plotted. Statistical significance greater than 1 (no category selectivity) for each set of generated activations are shown (blue text).

Certainly, there are many details left out of such a model, yet the reliance on empirical data grounds the model such that there is a clear path toward adding those details (e.g., fine-grained temporal dynamics) in future work. As a counterexample to illustrate the utility of the empirical constraint principle of activity flow modeling, consider the immense flexibility of modeling category-selectivity visual responses without such a constraint. As discussed further below, there are an enormous number of ways to generate a category selective pattern from exemplars (e.g., any of a variety of machine learning models or artificial neural networks (Cybenko, 1989)), such that we could spend decades demonstrating this variety of possibilities without narrowing down to a small set of empirically plausible options. The empirical constraint principle can help more rapidly narrow down the possibilities, reducing the scientific search space to those options consistent with both empirical connectivity and empirical activity patterns. Ideally, all relevant empirical constraints (e.g., the activity and connectivity of every neuron in the human brain) would be available to inform activity flow model-based explanations, but the work so far has demonstrated that we can still make tremendous progress without every relevant constraint. Indeed, the simplicity principle pushes against such a complex model, such that even if we had access to every neural event in the brain we would want to use simulated lesioning (Ito et al., 2022; R. D. Mill et al., 2022) or related dimensionality reduction methods to simplify the model and identify the core set of activity flows underlying functions of interest.

# Examples of successful activity flow models

We originally developed activity flow modeling to make two specific kinds of inference regarding neurocognitive phenomena (Cole et al., 2016). First, we wanted to determine where the human brain is on the continuum from primary localized (within-region) versus primarily distributed (inter-region) processing across a wide variety of tasks. Simulations were used to validate this inference – activity flow mapping only worked to the extent that underlying processes were distributed. The finding that activity flow mapping also performed well with empirical fMRI data suggested that the human brain's task-evoked activations are mostly determined by highly distributed processes. The second inference was to determine the functional relevance of resting-state FC to task-related processing. Prior studies had already shown a statistical link between resting-state FC and task-evoked activations (Smith et al., 2009), yet activity flow mapping attributed this to a common cause – that the activity flow pathways detected by resting-state FC were also those involved in generating task-evoked activations. Further, we developed an FC method that used multiple regression to control for (observed) causal confounds, likely resulting in a more causally-valid FC method relative to field-standard Pearson-correlation FC (see Sanchez-Romero & Cole, 2021). These (types 1 and 6) and other categories of activity-flow-based explanations are listed in **Table 2**, with the explanation types linked to individual studies in **Table 3**.

**Table 2 – The major forms of activity-flow-based explanations of neurocognitive phenomena (so far)**

| Type | What is being explained? | What is explanation based on? |
|---|---|---|
| 1 | Task-evoked activations | Distributed processes (activity flows) |
| 2 | Task-evoked activations | Specific connections and/or (other) activations |
| 3 | Dysfunctions/differences (in activations) | Specific connections and/or activations |
| 4 | Behavior (via motor activations) | Specific connections and/or activations |
| 5 | Task-related information (MVPA) | Specific connections and/or activations |
| 6 | Functional relevance of connection/activation | Contribution of connection/activation to another activation/dysfunction/behavior |

**Table 3 – List of studies that use activity flow modeling, what kinds of explanation they include, what kind of connectivity they use, and what kind of activity they use.**
DWI = diffusion weighted imaging. EEG = electroencephalography. RSFC = resting-state functional connectivity. Multreg = multiple regression. PCR = principal components regression. MVAR = multivariate autoregression (related to Granger causality). PC-algorithm = Peter-Clark algorithm. CombinedFC = combined functional connectivity (from (Sanchez-Romero & Cole, 2021)).

| Study | Explanation type | Imaging method | Connectivity type | Activity type |
|---|---|---|---|---|
| Cole et al. (2016) | 1,6 | fMRI | RSFC with correlation, Multreg, PCR | 7 diverse tasks, whole cortex; simulations |
| Ito et al. (2017) | 5 | fMRI | RSFC with PCR | 12 task rules, whole cortex; simulations |
| Mill et al. (2020) | 3,6 | fMRI | RSFC with PCR | Aging-related dysfunctional task-evoked activations |
| Ito et al. (2020) | 1 | fMRI | RSFC with Multreg | Activity in transmodal (vs. sensory-motor) cortex is better predicted by distributed processes |
| Cole et al. (2021) | 2,6 | fMRI | Task-state FC with correlation, Multreg, PCR | 24 diverse task conditions, whole cortex |
| Keane et al. (2021) | 2 | fMRI | RSFC with Multreg | Visual shape completion task activations |

| Hearne et al. (2021) | 3, 2 | fMRI | Cross-task average FC with Multreg | Working-memory-related activation dysfunction in schizophrenia |
| Yan et al. (2021) | 6 | DWI, fMRI | DWI structural connectivity | 2 tasks, whole cortex |
| Schultz et al. (2022) | 1,2,5 | fMRI | RSFC with correlation | 12 task rules, multiple demand regions |
| McCormick et al. (2022) | 6 | fMRI | Latent FC (rest & task) with correlation | 24 diverse task conditions, whole cortex |
| Ito et al. (2022) | 4,2,6 | fMRI | RSFC with Multreg | 64 context-dependent tasks, from sensory & rule activations to motor responses (in M1) |
| Mill et al. (2022) | 4,2,5,6 | EEG | RSFC with MVAR and PCR | Decoded motor information from source-localized motor cortex EEG activity |
| Hwang et al. (2022) | 2,3 | fMRI | RSFC with correlation | 100+ task conditions, thalamus as source and cortex as target |
| Zhu et al. (2023) | 6 | fMRI | Probabilistic correlation RSFC | 7 diverse tasks, whole cortex |
| Keane et al. (2023) | 3,6 | fMRI | RSFC with Multreg | Dysfunctional visual shape completion task activation in schizophrenia |
| Sanchez-Romero et al. (2023) | 6,2 | fMRI | RSFC with correlation, Multreg, combinedFC, and PC-algorithm | 24 task conditions, whole cortex; prefrontal working memory activation; simulations |
| Cocuzza et al. (Preprint) | 1,2 | fMRI | RSFC with combinedFC | 4 visual category activations, visual-category-selective regions |

Explanation type 2 (**Table 2**) is a more targeted inference, wherein a specific connection or activation (or a set of connections or activations) help explain the generation of a task-evoked activation. For example, Hearne et al. (2021) calculated individual activity flow estimates – the [activations * connectivity] values (prior to summing to produce the target activation). These values were then used to infer the source activations and connections that drove working-memory-related activations. Further, this approach was extended to group differences (explanation type 3), to identify the likely sources of dysfunctional working memory activations in schizophrenia patients. Other explanation type 2 efforts have involved simulated lesions (R. D. Mill et al., 2022) or network subset analyses (Keane et al., 2023) to isolate the sources of task-evoked activations of interest. This approach has also been extended to characterize the flow of task-related information via multivariate pattern analysis (explanation type 5), revealing specific information flows between brain regions (Ito et al., 2017) and the role of resting-state FC in determining the multifunctionality of cognitive control regions (Schultz et al., 2022).

Behavior holds a privileged place in the demonstration of neurocognitive functionality (Krakauer et al., 2017). We have therefore sought ways to link activity flow processes to behavior (explanation type 4). Based on our mechanistic/causal principle (**Table 1**), we determined that proper modeling of behavior would involve identifying the activity flows that generate motor responses via primary motor cortex. In Ito et al. (2022) this involved linking sensory inputs (visual and auditory) to task-rule-related activations via resting-state FC, ultimately resulting in generated activations in primary motor cortex (M1) (**Figure 1C**). These M1 activations were then decoded to identify which button would have been pressed given each M1 activation pattern. This resulted in above-chance performance of a complex context-dependent cognitive task. The activity flow model that generated this behavior could then be analyzed for insights into the likely mechanisms that led to this non-trivial cognitive task performance. Motor behavior generation was also the goal in Mill et al. (2022) (**Figure 2C**), with the advantage of high temporal resolution to improve the directionality of causal inferences (via generating future brain activity).

## Assumptions of activity flow models

The core assumption of activity flow modeling is that activity flows – the movement/propagation of activity between neural populations – support neurocognitive computation. This assumption has so far not been problematic, however, given that the activity flow framework builds in a way to test this assumption via prediction of (typically already available) empirical activations. As emphasized in the above discussion of explanation type 1 above (**Table 2**), this contrasts with the alternative possibility that processes within neural populations (recurrent processes) are essential for neurocognitive computation (and can't be accounted for with simple shifts in connectivity weights or blurring of time). Note that the recently developed high-temporal-resolution version of activity flow mapping can incorporate local recurrent processing as well as cross-region activity flows (R. D. Mill et al., 2022), potentially allowing direct comparisons between distributed and local processes.

From a graph theoretical perspective, activity flow mapping has so far assumed that activity propagates in the brain via a flow-based or diffusion-based routing protocol (Avena-Koenigsberger et al., 2017). Flow-based routing involves each signal (here task-evoked activations) propagating along all available direct connections rather than (for example) the most efficient/shortest path to a target node. This graph theoretical characterization suggests that – relative to shortest paths – activity flow models have assumed brain network communication optimizes for high parallel processing and low information costs (connectivity weights alone determine routing rather than a central router), but with high metabolic costs (since unnecessary nodes are activated). Note that these high metabolic costs may be offset by the sparsity of activations (due to local inhibition) (Rozell et al., 2008) and sparse connectivity (Sacramento et al., 2015). It will be important for future work to test this assumption in activity flow models, while also testing the optimization trade-offs predicted by graph theory.

When using FC, activity flow models assume that observed correlations (or other statistical associations) between neural populations are driven by similar activity flows as occur between task-evoked activations. For example, with resting-state FC this assumes that spontaneous activity flows at rest use the same (or similar) pathways as task-evoked activity flows. For task-state FC (with task-evoked activity confounds removed (Cole et al., 2019)) this assumes that spontaneous activity flows (or activity flow variation driven by trial-by-trial variability in stimuli) during a given task use the same pathways as task-evoked activity flows. Finally, when using non-directional FC this assumes that all connections are bidirectional and equally weighted. Note that weighted and directed connectivity in macaque monkeys suggests this tends to be the case (Markov et al., 2014). However, Sanchez-Romero et al. (2023) demonstrated the utility of directed functional connectivity (sometimes termed effective connectivity) approaches for making directional activity flow inferences with fMRI data, while Mill et al. (2022) did so with EEG data. Overall, the assumption that the same pathways are involved in generating the data used for estimating FC and the data used for estimating activations is not a problematic assumption, since activity flow predictions are unlikely to be accurate if this were not true.

Structural connectivity-based activity flow models (Yan et al., 2021) assume that activity flows occur over structural connections, which is almost certainly true. Such models also assume that all structural connections are bidirectional and symmetrically weighted, while the aggregate effect of synaptic weights (and perhaps other functional details captured by FC estimates) are negligible on activity flows. This is not a problematic assumption in the sense that activity flow-based predictions are unlikely to be accurate if this is not true.

An important assumption regarding the data used in activity flow models is the statistical independence of the connectivity and activity estimates. As covered in the "Pitfalls of activity flow modeling" section below, there are several ways that such statistical circularity can bias activity-flow-based inferences. For instance, if task fMRI data is used for both FC and activation estimation it is likely that activity-flow-based predictions would be biased toward overly accurate predictions of "held out" activations. This is due to the same variance (e.g., large task-evoked activations in two regions) driving both increased FC estimates and increase task-evoked activation estimates. This particular case is due to causal confounding (Figure 2) from external stimuli, such that regression of stimulus timing can remove this issue in some cases (Cole et al., 2019), making it possible to use task-state FC with activity flow modeling (Cole et al., 2021).

Most activity flow models have so far assumed stationarity of both FC and task-evoked activations, given their focus on time-averaged effects. Thus, most activity flow models will not capture changes to activity flows based on FC changes over time or task-evoked activations varying from trial-to-trial. Note, however, that some activity flow dynamics driven by transient task-induced activity are captured in the EEG activity flow modeling approach (Figure 3). It will be important for future studies to go beyond time-averaged effects to better characterize activity flow dynamics.

## Building, testing, interpreting activity flow models

From a broad perspective, activity flow mapping involves building a connectionist artificial neural network from empirical connectivity estimates. In principle, substantial biological detail can be added to these models, but the immense functional expressiveness of connectionist-level modeling (Rogers & McClelland, 2014) suggests that more can be learned by abstracting away from much of that detail when possible. This reveals a tension between the mechanistic and simplicity principles (**Table 1**), yet the optimal outcome from this tension should be the minimum mechanistic details necessary to explain/generate a given function of interest. Once this connectionist model is derived from empirical data, task-evoked activations are added, followed by the propagation and activation rules from standard connectionist modeling (Ito et al., 2020). This "animates" the model, simulating functions of interest and allowing inferences regarding the neural mechanisms underlying those functions to the extent that the model's constraints are empirically valid, the function is accurately generated by the model, and subsequent analyses isolate key factors within the model producing the function of interest.

More practically, one can build an activity flow model whenever one has both connectivity and activity estimates for a set of neural nodes (neurons or neural populations). Building and evaluating an activity flow model involves 6 steps. For visual overviews of how to build and test activity flow models, see Figure 1 for slow imaging methods (e.g., fMRI) and Figure 3 for fast imaging/recording methods (e.g., EEG). Step 1 is estimating connectivity among all nodes, such as using multiple regression (Cole et al., 2016). Step 2 is to estimate activity in all nodes, such as using an fMRI general linear model to estimate average task-evoked activations across trials of each condition. Alternatively, it is possible to use activity time series with activity flow modeling (see Figure 3), rather than averaging activity over time. Step 3 is to decide, based on the goals of the study, which nodes are *source nodes* and which are *target nodes*. Source nodes' activity levels are set to match empirical levels of activity in those nodes, while target nodes' activity levels are generated by the model. Step 4 involves running the model to generate activity in the target nodes across all task conditions of interest. Step 5 is to evaluate model accuracy by comparing generated to empirical activity in the target nodes. This can be done by a variety of similarity measures, such as correlation, $R^2$, or mean squared error. Assuming the generated activity matches empirical reality, step 6 involves gaining additional insights into the generative activity flow processes by interpreting the model features and intermediate activity (and activity flows) generated by the model. This can involve description of model features or interventions on model features (e.g., connections or activity levels) to observe the impact of those interventions on the generated activity.

For a detailed overview of how to implement activity flow modeling using the open source Brain Activity Flow Toolbox please see Cocuzza et al. (2022) and the toolbox website: https://colelab.github.io/ActflowToolbox/. Activity flow mapping was originally developed to relate standard task-evoked activations and standard FC as estimated using fMRI (Cole et al. 2016). Thus, these standard measures can be used to make activity-flow-based inferences (see Table 2). When optimizing for the theoretical principles underlying activity flow modeling (Table 1), however, other activity and connectivity estimates may be desired. The Brain Activity Flow Toolbox is highly flexible, allowing use of a wide variety of possible activity and connectivity types. Note, however, that activity flow mapping has only been adapted to high temporal resolution data (in this case source-localized EEG) recently. Until this is incorporated into the full toolbox, this adaptation of the approach can be found here: https://github.com/ColeLab/DynamicSensoryMotorEGI_release.

As a further guide to conducting studies using activity flow modeling, a list of steps are included below based on a set of recent activity flow modeling studies from my lab (Hearne et al., 2021; Ito et al., 2022; R. D. Mill et al., 2022; Sanchez-Romero et al., 2023), which focus on identifying connectivity-based explanations of specific neurocognitive phenomena:

1) Identify a brain function of interest, and obtain a reliable empirical measurement of that function. Examples: cortical hemispheric lateralization during language tasks; face selectivity in FFA (Figure 4); motor selection and responding in M1 (Figure 3).
2) Work backward from the function of interest to hypothesized source regions/times, excluding the to-be-generated function from the source set. Note that it is possible to remove the function via artificial means, such as averaging across hemispheres for a lateralization-generation study. For example, averaging the source task-evoked activations across hemispheres would allow the activity flow model to demonstrate that the model's network architecture was sufficient for generating lateralization, rather than simply spreading pre-existing lateralization to held-out brain regions. As a best practice, it is usually important to run an analysis verifying that the function of interest is not present in the source set, such that a connectivity-based transformation is required to generate that function in the model.
3) Use activity flow modeling to generate the function of interest based on the source activity flowing over connectivity, testing for above-chance predicted-to-actual similarity.
4) Identify the key connectivity (and/or activity) properties that allowed for the generation of the function of interest. Some example approaches include simulated lesions (Ito et al., 2022; see R. D. Mill et al., 2022), plotting/analyzing the activity flow graph (see Hearne et al., 2021), scrambling/permuting connectivity patterns to show dependence on those specific patterns, and graph theory to identify connectivity properties (ideally comparing to a model with those properties removed, or identifying correlations between the strength of those properties and variation in the function of interest).

# Pitfalls of activity flow models

Perhaps the most commonly encountered pitfall in activity flow modeling is the possibility of analysis circularity (Kriegeskorte et al., 2009) in the comparison between generated and actual task-evoked activations. This typically arises from mixing source and target node activity during activity flow simulation. This leads to some portion of the to-be-predicted (target) data being erroneously added to the model-produced activity, rather than having the model generate that activity independently. For example, the spatial smoothness inherent in fMRI data creates circularity in activity flow inferences between nearby voxels. Cole et al. (2016) dealt with this issue by using a brain region atlas with regions 10 mm apart and, when performing voxelwise analyses, excluding all voxels within 10 mm of each target voxel from the set of source flows. This has become common practice in activity flow mapping with fMRI data. Note, however, that tests of the impact of circularity with brain region atlases (using cross-voxel averaged time series) that share borders have revealed that there are typically only minimal impacts of circularity in these cases (see Activity Flow Toolbox demo: https://colelab.github.io/ActflowToolbox/HCP_example.html). Nonetheless, it is a best practice to avoid source-target pairs that are within 5 mm of each other (and likely more with voxels larger than 2.5 mm) with fMRI data.

The situation would appear to be much worse with EEG or MEG data, as volume conduction can cause a source signal to traverse the entire brain, causing analysis circularity for every possible source-target pair. Mill et al. (2022) used a rigorous multipronged approach to eliminate this issue, however. Most consequentially, simulated activity flows generated *future* activity to completely avoid the possibility that volume conduction (which occurs with zero lag) could result in analysis circularity. Additional steps increased the precision of activity flow inferences. First, high-density EEG data and structural MRIs were used to localize sources with beamforming. Second, sources were localized to the same set of regions with 10 mm gaps used by Cole et al. (2016), reducing the chance of substantially mis-localized sources. Third, rather than using standard temporal filters we used causal filters, which prevent temporal circularity by avoiding leakage of future activity back in time. Finally, the target region's zero-lag time series was regressed out of all source time series, while also being fit in an autoregressive manner, further improving isolation of the target time series. Together, these steps both eliminate the chance of spatial analysis circularity while substantially improving inferential precision.

Another source of potential circularity is specific to using task-state FC (rather than resting-state FC or structural connectivity): Simultaneous sensory inputs leading to inflation of FC and inflating activity flow prediction accuracy (Cole et al., 2019). For example, watching a video of a car exploding will simultaneously stimulate your primary auditory and visual cortices (A1 and V1), creating strong correlations between those regions despite there being no direct connectivity between them. This situation would create a false functional connection between A1 and V1, which could then be used with an activity flow model to predict V1 activity based on A1 activity (and vice versa). The resulting misleading inference would be due to a causal confound, wherein an external common cause has created the appearance of direct causal influence for both the FC estimation algorithm and the activity flow model. Cole et al. (2019) identified this confound and demonstrated that subtraction (or regressing out) of cross-trial mean task-evoked activity is a way to remove the impact of the common cause and correct for this confound. This is therefore the best practice when using activity flow modeling with task-state FC (Cole et al., 2021).

A general pitfall of activity flow modeling is shared with any connectivity-based study: the chance that inaccurate connectivity estimates will lead to an inaccurate activity flow-based inference. Multiple studies have demonstrated that improving FC estimation also improves activity flow-based task-evoked activation prediction accuracies (Cole et al., 2016; Sanchez-Romero et al., 2023). We define FC improvement as improved estimation of cross-neural-population causal interactions (Reid et al., 2019), such as reducing the number of causal confounds. While prediction accuracy typically increases with improved causal FC (since the true causal model is among the best possible predictive models), it remains possible for prediction accuracy to be improved by causal confounds (Reid et al., 2019; Sanchez-Romero et al., 2023). The example above with task-state FC confounds is a case in point – the causal confound of correlated inputs can enhance prediction accuracy of A1 activity from V1 (and vice versa). Therefore, it is essential to rely on bedrock causal principles – rather than activity flow prediction accuracy alone – when developing connectivity methods for parameterizing activity flow models (Sanchez-Romero & Cole, 2021; Sanchez-Romero et al., 2023).

A non-obvious case of analysis circularity can occur when a causal inference is in the wrong direction. For example, it is known that primary motor cortex sends efferent copies of its output back to brain regions that are inputs to primary motor cortex (Khan & Hofer, 2018). This is less of a problem with high-temporal-resolution methods like EEG, since the feedforward and feedback processes can be easily separated using dynamic activity flow mapping (R. D. Mill et al., 2022). Separation of feedforward and feedback processes is much more challenging with a low-temporal-resolution method like fMRI, however, since the feedforward and feedback processes (which likely occur on the order of 100-200 ms) are mixed into single fMRI task-evoked activation estimates. There are three basic strategies (so far) to deal with this issue:

First, simply make a weaker inference, wherein source activations caused target activations *and/or* target activations caused source activations. Cole et al. (2016) ran simulations to show that even this weaker inference is useful, as it can reveal the extent to which distributed processes support the generation of task-evoked activations (explanation type 1 in **Table 2**).

Second, use multi-step activity flow modeling starting from sensory inputs, wherein only activation patterns in primary sensory regions (e.g., V1) are empirical and all subsequent activity flows are based on that input (Cocuzza et al., 2022). This isolates variance to feedforward processes from the inputs, eliminating the chance for causal circularity after that initial step. There is some chance of feedback processes impacting the initial input node, however, resulting in causal circularity in the initial activity flow step. Cocuzza et al. (2022) dealt with this issue via signal normalization to ensure that there was no selectivity for each visual category of interest in V1 (the input node). In other words, the input node did not contain the function of interest, ensuring that the function of interest was ultimately generated via transformations on the input node's activity.

Third, use of careful counterbalancing and averaging to remove the to-be-generated effects of interest from the source task-evoked activations, such that feedback processing cannot explain observed effects. For example, Ito et al. (2022) use of a counterbalanced factorial design, wherein all stimuli were paired with each task rule and each motor response across trials. They then averaged across all task rules and motor responses when estimating each stimulus input activation. The resulting counterbalanced-and-averaged task-evoked activation estimates guaranteed that motor responses generated via mixing of stimulus and task rule activity flows would occur in a non-circular manner.

Together, these strategies reveal the flexibility of the activity flow modeling framework. Even in a situation where causal circularity seemed inevitable, clever use of activity flow algorithm development (e.g., a multi-step approach) and experimental design (e.g., counterbalancing) led to inferential improvements. I expect future innovations to overcome current and future challenges to the activity flow modeling framework.

# How activity flow modeling relates to other approaches

Activity flow modeling is highly related to a variety of other approaches to brain data analysis and neural network modeling. Despite being similar to these approaches, activity flow modeling does not appear to be redundant with any of them, and can add something unique to each of them. Indeed, it appears that any possible connectivity method – and any possible activity estimation method – can be incorporated into activity flow models, with unique inferences added in each case. Further, activity flow modeling can add unique inferences to standard task-evoked neural activation studies as well as theoretical computational models. These points are illustrated below. **Figure 5** provides a general map along mechanistic and empirical axes, illustrating the relationship to activity flow modeling to some other approaches.

|  | Less Empirical | More Empirical |
|---|---|---|
| More Mechanistic | Biologically detailed theoretical models<br><br>Artificial neural networks (e.g., deep learning) | **Activity flow** (V1-initiated; Cocuzza et al. 2022)<br><br>**Activity flow** (fMRI; Cole et al., 2016) |
| Less Mechanistic | Box-and-arrow cognitive models | Encoding models (predicting neural activity from stimulus/ task condition)<br><br>Predicting individual differences in behavior/cognition from brain data (e.g., resting-state FC) |

**Figure 5 – Relationship between activity flow modeling and some other approaches to modeling neurocognitive phenomena (behavior and task-evoked activations).**
A subset of neurocognitive modeling approaches are shown along two axes: mechanistic and empirical. The placement of the text describing each approach indicates approximately (even within each quadrant) its relative level of mechanistic vs. empirical properties. *Mechanistic* is defined here as the level of detail regarding the generative causal processes hypothesized to underly the phenomena of interest. For example, box-and-arrow cognitive models (e.g., Baddeley, 2000) include fewer details regarding the causal events hypothesized to generate cognitive processes than connectionist artificial neural networks, and even fewer than biologically detailed theoretical models (e.g., Babadi & Abbott, 2010). Activity flow modeling was developed to be more empirically constrained and empirically validated than artificial neural network models, and more mechanistic than models that predict behavior (or task-evoked activations) based on brain data (e.g., Smith et al., 2009, 2015). Two activity flow models are shown, with the Cocuzza et al. (2022) V1-initiated model (Figure 4B) being more mechanistic than the original activity flow modeling approach (Cole et al. (2016)). This is because of the additional details included regarding the generative causal processes underlying the generation of the processes of interest (in this case visual category selectivity in human visual cortex). Note that so far no activity flow model has been as mechanistically detailed as biologically detailed theoretical models of neural function (but see Lee et al., 2022).

*Computational models (deep neural networks, recurrent neural networks, etc.)*. Activity flow models can be considered to be computational models that – unlike standard computational models – are derived from empirical brain data. Indeed, activity flow models were developed based on connectionist artificial neural networks (ANNs), which consist of nodes and connections over which activity flows according to a standard "propagation rule" (Ito et al., 2020; Rogers & McClelland, 2014). Importantly, hundreds of studies over decades have revealed the seemingly limitless potential of ANNs to model human cognition (McClelland & Rogers, 2003; Rumelhart, Hinton, McClelland, et al., 1986; Rumelhart, Hinton, & Williams, 1986) and can even exceed human cognition in "deep neural network" versions of ANNs with many layers (Bengio et al., 2021; Silver et al., 2018; Vaswani et al., 2017). This flexibility of ANNs is both a blessing and a curse, as ANNs can be used to model anything (Hornik et al., 1989) yet they often fit the training data differently than the human brain does. Indeed, this divergence from the human brain is highly likely, given the immense architectural differences between ANNs and the human brain. A major motivation for developing the activity flow framework was to add extensive empirical constraints in order to improve ANN-based inferences regarding the human brain and human behavior. Note that some prior studies already used empirical data as inputs to ANNs (Hanson & Hanson, 1996), illustrating the utility of empirical data constraints (as opposed to broad empirical constraints, such as use of activity propagation among units to model brain processing) on neural network-based inferences. Further, given that humans have computational/cognitive abilities that ANNs do not, there is potential to develop computational models with state-of-the-art abilities based on brain data-derived modeling. In principle, using activity flow modeling to create an ANN architecture directly from empirical brain connections should 1) provide empirical neural data analyses with the theoretical insights typical of computational modeling and, 2) provide computational modeling with the scientific conclusiveness and grounding in reality of empirical neural data analyses.

Some recent deep ANNs have taken inspiration from the network architecture of the primate visual system, roughly matching the number of layers/steps used by primates to process visual stimuli (see Yamins & DiCarlo, 2016 for review; Yamins et al., 2014). These studies have shown that – despite primarily being shaped by learning an object recognition task – each ANN layer's representations are similar to the kinds of representations present in empirical multi-unit recording data from non-human primate brains. Together, these results demonstrate that even rough approximations of the correct network architecture can (when augmented by connectivity adjustments from task learning objectives) result in important insights into neural computations. Activity flow modeling goes beyond these models by using network architectures *directly specified by brain data* (rather than task learning objectives), incorporating many more brain network architectural features and thus providing results that are likely to be even more informative regarding the network-based cognitive computations carried out by the brain.

*Resting-state FC (with fMRI, magnetoencephalography (MEG)/EEG, intracranial EEG, etc.).* Activity flow modeling has so far been primarily used to add insight into the computational and cognitive contributions of resting-state FC (Cole et al., 2016; Ito et al., 2020). This complements other approaches that associate resting-state FC with cognition, such as correlating resting-state FC with activity patterns (Smith et al., 2009) or individual differences in behavior (Cole et al., 2012; Smith et al., 2015). This is accomplished by parameterizing activity flow models using resting-state FC estimates and a subset of task-evoked activations, then testing for the resulting model's ability to generate held-out task-evoked activations linked to cognitive phenomena. High generated-to-actual similarity (compared, e.g., to randomly permuted models) provides evidence for the validity of the activity flow model, and thus supports the associated model-based explanation for the generation of task-evoked activations and associated cognitive phenomena. Notably, the mechanistic principle described above further increases confidence in the validity of the model-generated explanation, since it provides additional constraints on the model beyond parameterization with empirical connectivity and activity (and testing of the model via comparison with empirical data). As a case in point, we typically use multiple regression (rather than field-standard Pearson correlation) to estimate FC for fMRI activity flow predictions (Cole et al., 2016, 2021), given the improved mechanistic/causal validity of multiple-regression FC due to reduced FC confounds (**Figure 2**).

A major reason to use resting-state FC (as opposed to FC during any given task state) is the desire to identify FC architectures that generalize beyond specific states. This idea of identifying a general network architecture has resulted in the development of latent FC (McCormick et al., 2022), wherein factor analysis is used to identify the FC weight underlying a wide variety of task (and rest) brain states. Even without using latent FC (which, in its standard form, requires each subject to perform a task battery), this concept can be used to strengthen the generalizability of activity flow inferences by simply estimating FC with an independent brain state (e.g., rest, or a different task) from the task state(s) of interest. For example, Hearne et al. (2021) averaged FC across rest and a variety of tasks, excluding the task of interest from FC estimation to reduce the chance of analysis circularity and help ensure the results were based on a state-general brain network architecture.

*Individual differences resting-state FC.* As briefly described above, a variety of approaches use resting-state FC to predict individual differences in cognition and behavior (Cole et al., 2012; Shen et al., 2017; Smith et al., 2015; Varoquaux & Poldrack, 2018). Further, some recent approaches use resting-state FC and/or structural connectivity to predict individual differences in task-evoked activations (Bernstein-Eliav & Tavor, 2022; Osher et al., 2016; Tavor et al., 2016). Like these approaches, the activity flow modeling approach is predictive. However, rather than predicting data from held-out individuals, the activity flow approach predicts held-out activations *within individuals*. While prediction can be implemented just as well between individuals, causal brain mechanisms ultimately occur within individual brains and are thus better served via within-individual models using causally grounded connectivity estimates. As an illustration of this point, consider how challenging it would be to use an individual differences approach to infer the functions of the human heart and circulatory system, given how little the relevant mechanisms differ between individuals (resulting in very small individual difference correlations) and how the relevant biological mechanisms all interact in complex ways within individual bodies. Thus, activity flow modeling differs from these other methods in its primarily mechanistic and causal explanatory (rather than primarily predictive) goal (see Table 1).

*Task-state FC (with fMRI, MEG/EEG, intracranial EEG, etc.).* Task-state FC (also termed task-related FC) – FC estimated from brain data collected during task performance – can be used nearly as easily as resting-state FC to make inferences with activity flow modeling. This was recently demonstrated by Cole et al. (2021), wherein task-state FC was shown to consistently improve activity flow predictions beyond those obtained using resting-state FC. This suggests that task-state FC estimated from the same task as the to-be-predicted task activations contains additional features that determine the flow of activity above and beyond those present in other states such as rest. This makes an entirely new line of research possible, wherein task-state FC changes specific to a given task condition are identified and their functional relevance (for generating task-evoked activations) determined via activity flow modeling. Note, however, the need to account for potential analysis circularity when using task-state FC (covered below).

*Structural connectivity (with diffusion weighted imaging (DWI), tract tracing, etc.).* A recent study demonstrated that DWI-based structural connectivity can be used effectively in place of resting-state FC for activity flow modeling (Yan et al., 2021). The theoretical inferences possible with structural connectivity are very similar to those of resting-state FC. Indeed, one can think of resting-state FC as structural connectivity with additional information regarding the aggregate effects of synaptic weights. However, relative to correlation-based FC, structural connections are much more likely to reflect direct connections between neural populations (Damoiseaux & Greicius, 2009). (Note that other FC methods – such as regularized partial correlation – can also be used to estimate direct connectivity.) Further, unlike FC estimates (and some forms of tract tracing), DWI-based structural connectivity has a strong distance-based bias toward false negative connectivity (Donahue et al., 2016). Thus, there are some advantages and some disadvantages to using structural connectivity relative to FC for activity flow modeling.

*Task-evoked activation-based approaches (GLMs, MEG/EEG event-related potentials, multivariate pattern analysis (MVPA), etc.).* Task-evoked activations – changes in neural activity amplitude – are both the inputs and the to-be-generated outputs of activity flow models. This reflects the principle of empirical grounding (see **Table 1**), wherein we maximize contact with empirical data by using empirical activations as input into each model. Importantly, our

explanatory inferences also focus on activations, given the mechanistic centrality of activations in standard neural theory. Specifically, the construct of a neural action potential entails both activation (i.e., an increase in neural activity amplitude) and the flow of activity to other neurons. Neural activity amplitudes are thus ultimately based in action potentials/spike rates, which cause local field potentials, large-scale electromagnetic fields, synaptic activity, hemodynamic responses (Lee et al., 2010), and other measures of neural activity. Thus, activity flow modeling leverages standard, well-established neural theory to interpret task-evoked activations and their flow/movement to other neural populations via connectivity patterns. These inferences are then verified empirically by comparing the generated task-evoked activations to the actual task-evoked activations in each neural population for each task condition (**Figure 1B**). This theoretical background reveals that the ideal activation estimation would use multi-unit spike recording. In the absence of such ideal data in humans, we use task GLMs with fMRI (**Figure 1 & Figure 4**), event-related potentials with MEG/EEG (**Figure 3**), or any of a number of possible activation-derived functional/cognitive signatures.

Activity flow mapping has also been applied to gain insights into the flow of neural information content via modeling multivariate patterns of task-evoked activations (Ito et al., 2017). To illustrate one possible approach, consider using MVPA to decode information content in each brain region, then using the activity flow approach to predict (based on other regions' information content) downstream information content based on the connectivity strength between the regions. In the interest of maintaining the mechanistic principle (see **Table 1**) to improve our inferences, however, we used a more nuanced approach (but see Schultz et al., 2022 for a simpler approach). This involved first applying standard activity flow mapping between individual vertices (the smallest unit of spatial measurement with cortical surface fMRI data). All vertex-to-vertex resting-state functional connections between each pair of cortical regions were estimated. Empirical task-evoked activity for each vertex in the source region was used as input, then the activity flow calculation was applied to predict task-evoked activation patterns in the target region. MVPA was then applied to this predicted activity pattern, based on a decoder trained to distinguish the actual activity patterns across task conditions. This procedure allowed us to infer the degree to which brain regions shared task-related information via fine-grained resting-state FC topology. More generally, this study demonstrated that activity flow mapping can be readily applied to make inferences about multivariate patterns of activity (and connectivity), suggesting the ability to extend activity flow mapping to a variety of research questions relating to neural information representation. For instance, this study was the basis of several recent studies that incorporated inferences regarding the flow of task-evoked information-carrying activity patterns (Hwang et al., 2022; Ito et al., 2022; R. D. Mill et al., 2022; Schultz et al., 2022).

*Encoding models*. As discussed by Ito et al. (2020), activity flow models can be considered to be connectivity-based encoding models. This contrasts with standard encoding models (Huth et al., 2012; Naselaris et al., 2011) (**Figure 6A**), which predict task-evoked activations directly from stimuli (or task conditions) mapped (via a regression weight) to a given neural population (e.g., an fMRI voxel). Note that standard task fMRI GLMs can be considered encoding models, though they are not typically tested as predictive models using properly held-out data. Activity flow models also predict task-evoked activations but, rather than using a direct mapping from stimulus features (or task conditions) to task-evoked activations, brain connectivity is used to map task-evoked activations in one or more neural populations to task-evoked activations in another neural population (**Figure 6C & 6D**). Critically, this connectivity-based approach is more mechanistic than standard encoding models (**Figure 5**), since connectivity is thought to be a primary mechanism by which task-evoked activations are generated in the brain (Cole et al., 2016; Ito et al., 2020; Mars et al., 2018; Passingham et al., 2002). Thus, even if an activity flow model makes less accurate predictions of task-evoked activations than a standard encoding model, that activity flow model will still provide insight due its additional connectivity-based mechanistic details. The Ito et al. (2022) study is the most complete example to date of the full stimulus-to-response activity flow modeling approach illustrated in **Figure 6C**. That study (**Figure 1C & 6D**) started with standard encoding models (visual, auditory, and task set activations) and combined them with empirical connectivity patterns to generate cognitive representations. Those cognitive representations then generated motor representations implementing context-dependent cognitive task performance (as verified by a decoding model).

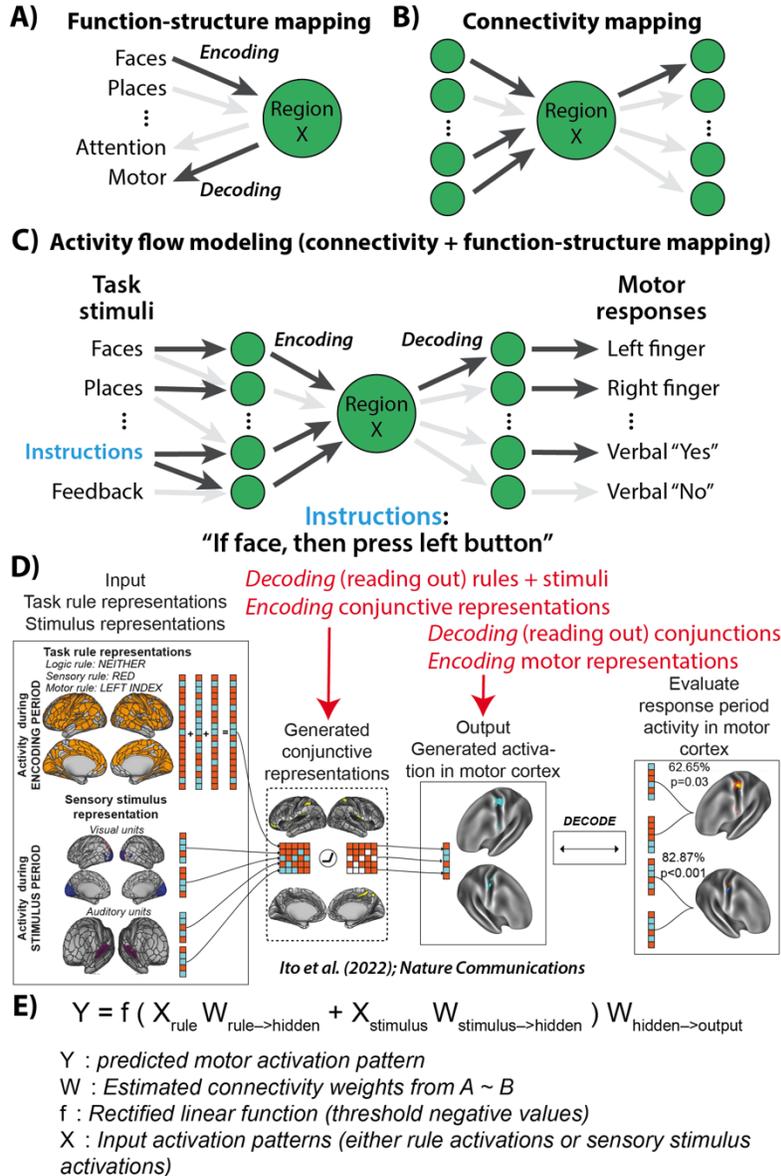

**Figure 6 – From encoding and decoding to activity flow modeling.**
**A**) Function-structure mapping (Henson, 2005), such as standard encoding models (e.g., fMRI GLMs) and decoding models (Naselaris et al., 2011). **B**) Connectivity mapping, such as using diffusion MRI to estimate structural connectivity or using fMRI to estimate functional connectivity. **C**) Activity flow modeling combines connectivity mapping with function-structure mapping, inferring how functions are generated via activity flows over connectivity patterns. Rather than the experimenter doing the encoding/decoding, activity flow modeling allows us to infer how neural populations decode each other (de-Wit et al., 2016) to encode information when implementing neurocognitive processes. Panels A-C adapted from Ito et al.

(2020). **D**) Results from Ito et al. (2022) – an example of a full activity flow model mapping representational transformations all the way from stimulus inputs to motor outputs in a context-dependent cognitive task. Each intermediate step can be interpreted as both a decoding model (decoding input activity patterns) and an encoding model (encoding output activity patterns). Decoding is used as the final step to determine what behavior (i.e., motor actions) is generated by the model. The model generated above-chance task performance. **E**) Equation specifying motor (output layer) activity patterns in panel D, based on activity flowing between model layers. Note that the "hidden" layer is labeled as "Generated conjunctive representations" in panel D.

## Open challenges and future directions

The success of activity flow modeling depends on the quality of the methods used to estimate the task-evoked activations and connectivity that parameterize activity flow models. Thus, improvements to connectivity and task-evoked activation estimation are inevitably also improvements to activity flow mapping. For example, we recently made the case for improving FC approaches by shifting the goal of FC research from estimating associations to estimating causal relationships (Reid et al., 2019). Several of our recent studies have supported improved causal inferences with improved FC measures (Cole et al., 2016; Sanchez-Romero & Cole, 2021), resulting in improved activity flow inferences (Sanchez-Romero et al., 2023).

Switching to a causal goal for FC also opens up new means to validate (and improve) FC methods using causal stimulation approaches. For example, combining a stimulation approach like transcranial magnetic stimulation (TMS) and a neuroimaging approach like EEG can yield causally grounded connectivity maps (Esposito et al., 2020). This is based on evidence that TMS causes localized action potentials (Mueller et al., 2014), along with the assumption that this results in stimulation-evoked activity in brain regions downstream of the stimulated site. Importantly, this assumption is based on the activity flow construct – movement of experimenter-evoked activity between neural populations. Thus, activity flows are theoretically central to this general approach, with activity flow modeling empirical testing this via assessment of whether actual causal interventions (stimulations) applied to brain regions aligns with activity flow-generated effects of such interventions. Some alternative combinations of causal stimulation and brain recording that can support this FC validation approach include transcranial electric stimulation and fMRI (Kar et al., 2020), intracranial stimulation and fMRI (Thompson et al., 2020), and intracranial stimulation and recording (Sheth et al., 2022). Given the centrality of activity flows to these simultaneous stimulation-recording approaches, these approaches have the potential to both benefit and benefit from the activity flow framework. Indeed, activity flow modeling has strong potential for providing individualized predictions of stimulation interventions to reduce symptoms (i.e., treatments) across a wide variety of brain disorders (Sanchez-Romero et al., 2023).

Activity flows are directly tied to action potentials, given that action potentials are the only known mechanism for long-distance neural signal propagation in the brain. This theoretical link supports the likely centrality of activity flows in neural processing. There remains an important open challenge to confirm this link, however, given the implication that measures like fMRI task-evoked activations and FC could be more directly tied to action potentials (spike rates) with a simple mathematical transformation (i.e., estimated activity flows). As Carl Sagan said, extraordinary claims require extraordinary evidence. Such evidence in this case could come from simultaneous blood oxygen level dependent (BOLD) response measurements and multi-unit recording in brain region pairs, ideally with varying levels of connectivity among region pairs. The prediction that activity flow estimation brings fMRI/BOLD signals closer to spike rates would be supported if task-evoked activity flows better reflect spike rates than the BOLD task-evoked activations and FC estimates that went into those activity flow estimates.

An open challenge for the recently-developed high-temporal-resolution activity flow mapping approach (R. D. Mill et al., 2022) is the possibility of subtle analysis circularity driven by temporal autocorrelation. Specifically, it will be important to use activity flow to disambiguate between truly generating the onset of task-related information (information transformation) versus modeling its spread (information transfer) across the brain after that initial generation. For example, if a temporally-distributed task-evoked activation is the function of interest, the start of this activation may not be predicted, yet once the start of the activation becomes the input to the model it can accurately predict the rest of the activation's spatiotemporal pattern. In the end such a model would be just pattern completing the temporal profile of the activation rather than generating it. One potential solution is to define the entire activation as the function of interest, then ensure the inputs to the activity flow model do not include any portion of the activation itself, based on either spatial or temporal (or both) restriction of the inputs. This issue reveals how activity flow modeling touches on the deep philosophical question of emergence (Mediano et al., 2022). It will be important to clarify best practices surrounding the identification of neurocognitive phenomena of interest and what features can interact within activity flow models to properly explain their emergence.

Another open challenge for activity flow modeling is to expand it to more invasive methods, potentially improving causal inferences by measuring more direct neural signals relative to the noninvasive methods used so far. However, most invasive methods have a much smaller field of view than fMRI and MEG/EEG, reducing the chance of detecting and controlling for causal confounds when estimating FC (Pearl & Mackenzie, 2018; Reid et al., 2019). Thus, it is not necessarily the case that invasive methods will improve activity flow inferences. Nonetheless, the fundamental role of action potentials in neural processing, and the clear link between action potentials and activity flows, suggests that activity flow modeling can add insights into all manner of neuroscience data. Some of this potential has been realized in the form of a multi-region multi-unit recording study in non-human primates (Ito, 2021; Chapter 4). This study demonstrated the potential for activity flow mapping to link intrinsic FC and task-evoked activations even when using invasive spike recordings. Similarly, an intracranial EEG study was conducted with human participants with intractable epilepsy, revealing that high-temporal-resolution activity flow mapping is successful in modeling the generation of spectrally resolved language processes based on resting-state FC (R. Mill et al., 2022). Together, these studies point the way toward utilizing activity flow modeling across the wide variety of available neuroscience methods, extracting new insights from those methods and ultimately driving discovery of distributed mechanisms underlying brain functions.

# Appendix: How activity flow modeling relates to additional other approaches

*Dynamic causal modeling (DCM) (fMRI, MEG/EEG)*. DCM is a functional connectivity approach (often termed effective connectivity) that estimates causal interactions from neural time series by fitting highly parameterized models using a variational Bayesian estimation approach (Friston et al., 2003). In principle, DCM could be used with activity flow modeling just like any other FC measure. Indeed, DCM is highly compatible with the activity flow framework, given the emphasis on causal mechanisms and connectivity in both approaches. However, activity flow modeling goes further than DCM in characterizing the generative mechanism of neurocognitive processes of interest. This is due to the additional building of generative models – which could be based on DCM estimates in this case – focused on predicting each neurocognitive process of interest in independent data. This added value of activity flow modeling reflects the fact that the DCM framework does not demonstrate an ability to predict/generate neural effects in independent data. Additional inferential value can be added using recent activity flow modeling innovations, such as simulated lesioning to infer the importance of each model feature in generating the neurocognitive process of interest (Ito et al., 2022; R. D. Mill et al., 2022).

*Other complex neural modeling approaches.* A wide variety of complex neural (and cognitive) modeling approaches exist. In principle, activity flow modeling could be integrated with and complement any of them. For example, the "joint modeling of neural and behavior" approach (Palestro et al., 2018) identifies neural variance linked to behavioral variance, which could also be linked to activity flow processes as mediators between sensory input and motor output. A similar modeling framework (in that it encompasses both neural and behavioral data) is the dynamic neural field theory modeling framework (Wijeakumar et al., 2017). These models appear to be initially hand-built (based on researchers' assumptions/hypotheses), then researcher-specified free parameters are fit to neural and behavioral data. Activity flow modeling could be used to empirically specify/constrain the network architecture within each neural field theory model. This integration has the potential to improve inferences regarding how empirical brain connectivity specifies cognitive functionality in the human brain.

As another example, The Virtual Brain (Ritter et al., 2013) may initially appear to encompass activity flow modeling given that it generates activity time series based on (typically structural) connectivity. However, it appears the inferential algorithm underlying activity flow mapping – using empirical activity and connectivity to generate held-out task-related neurocognitive functions – is not a standard part of The Virtual Brain. Instead, The Virtual Brain uses complex biophysical simulation equations to model neural mass model dynamics, typically using randomly generated spontaneous activity as a starting point (rather than empirical activity). Thus, activity flow modeling could be readily incorporated into The Virtual Brain to expand the inferential power of the framework. One promising avenue for this incorporation of activity flow modeling is the ability to input task stimulus timing (or brain stimulation) time series into The Virtual Brain simulations. It appears this feature has not been widely utilized in published The Virtual Brain studies, as the primary utilization of this modeling framework has been to simulate resting-state brain dynamics via simulated dynamics over structural connectomes (Martí-Juan et al., 2023). More generally, it is worth noting that the simplicity/abstraction principle underlying the activity flow modeling framework (see Table 1) suggests the need to test The Virtual Brain's assumption that complex neural mass modeling equations are necessary to accurately generate neurocognitive functionality.

One important exception to the focus on resting state (rather than task state) processes with The Virtual Brain has been a series of studies incorporating a biologically realistic task-performing model (Tagamets & Horwitz, 1998) into The Virtual Brain modeling software (Liu et al., 2022; Ulloa & Horwitz, 2016, 2018). The primary advantages of this integration appears to be making more specific predictions regarding the anatomical locations corresponding to each of the original model's nodes, as well as more realistic generation of noise via The Virtual Brain's standard structural connectome and dynamical equations. Critically, however, the model's task-performing elements remain completely specified by the original version of the model, which was created via a mix of the researchers' domain knowledge and hand-coded connectivity (Tagamets & Horwitz, 1998). Thus, unlike activity flow modeling (Ito et al., 2022), it appears that task-related functionality has not been generated by lower-level activity and connectivity constraints within The Virtual Brain. This represents a major opportunity to increase the task-related cognitive relevance of The Virtual Brain simulations via incorporation of activity-flow-like modeling concepts.

*Feedforward predictive modeling of the visual system*. Activity flow modeling has recently been applied specifically to the visual system (Cocuzza et al., 2022), leveraging rich knowledge regarding visual processes in the brain to enhance activity flow models (and gain further insight into the visual system). Several other approaches have been developed based on similar insights regarding the role of activity flows in generating activity throughout the visual system. First, Haak et al. (2013) developed connective field modeling, which fits Gaussian spatial models that map visual receptive fields between cortical areas. This approach is directly related to the concept of activity flow over connectivity, but is more specifically related to topographic organization of representations within the visual system. It appears that the activity flow modeling approach could extend this approach to gain insight into the activity flow relationships among non-topographic features in the visual system. By the same token, connective field modeling could refine activity flow modeling to better reveal when activity flow projections are topographic in organization. A second approach, voxel-to-voxel predictive modeling (Mell et al., 2021), is again focused exclusively on the visual system, but is more directly related to activity flow modeling. Like activity flow modeling, relationships between neural populations (weights) are estimated, then source activity multiplied by the weights are used to predict target activity. However, rather than using connectivity per se, the voxel-to-voxel weights appear to be based on the task-evoked activities themselves. This results in a distinct set of inferences, wherein the role of connectivity in the predicted relationships is less clear. It appears that the voxel-to-voxel predictive modeling approach could be enhanced by the use of functional or structural connectivity to gain a more mechanistic understanding of the relationships between visual areas. Activity flow modeling could also be enhanced by taking insights from voxel-to-voxel predictive modeling, especially with regard to inferring receptive field properties, the relationship to stimulus encoding models, and comparing results to deep neural network models of the visual system.

## Take-home points

- Activity flow is the movement of activity between neural populations (also termed: activity propagation, information flow, activity spread, or activity diffusion)
- Activity flow modeling is a flexible approach that allows for building (based on empirical brain data) and empirically testing network models of brain function
- Task-evoked brain activity is used as input to simulate the flow of activity over empirically estimated brain connections, which is then tested for the ability to generate a neural or cognitive phenomenon of interest
- Examples of generated neurocognitive phenomena from previous activity flow mapping studies: face selectivity in the fusiform face area, task rule representations in the frontoparietal control network, and motor responses decoded from M1 activity (i.e., behavior) during a context-dependent decision-making task
- Four general principles guide activity flow modeling, together optimizing for clear and empirically grounded explanations of neurocognitive phenomena: 1) Simplicity (Occam's razor), 2) Generative, 3) Mechanistic/causal, and 4) Empirically constrained (data driven)
- Any connectivity method (and any activity estimation method) can be incorporated into activity flow models
- Activity flow modeling can add unique inferences to standard task-evoked neural activation studies as well as theoretical computational models

## Acknowledgements

I would like to thank Ravi Mill, Ruben Sanchez-Romero, Carrisa Cocuzza, Kirsten Peterson, Alexandros Tzalavras, and Lakshman Chakravarthy for their feedback on an earlier version of this chapter. This work was supported by US National Science Foundation grant 2219323.

# References

Avena-Koenigsberger, A., Misic, B., & Sporns, O. (2017). Communication dynamics in complex brain networks. *Nature Reviews. Neuroscience*, *19*(1), 17–33.

Babadi, B., & Abbott, L. F. (2010). Intrinsic stability of temporally shifted spike-timing dependent plasticity. *PLoS Computational Biology*, *6*(11), e1000961.

Baddeley, A. (2000). The episodic buffer: a new component of working memory? *Trends in Cognitive Sciences*, *4*(11), 417–423.

Bengio, Y., Lecun, Y., & Hinton, G. (2021). Deep learning for AI. *Communications of the ACM*, *64*(7), 58–65.

Bernstein-Eliav, M., & Tavor, I. (2022). The Prediction of Brain Activity from Connectivity: Advances and Applications. *The Neuroscientist: A Review Journal Bringing Neurobiology, Neurology and Psychiatry*, 10738584221130974.

Cocuzza, C. V., Ruben, S.-R., Ito, T., Mill, R. D., Keane, B. P., & Cole, M. W. (2022). Distributed resting-state network interactions linked to the generation of local visual category selectivity. In *bioRxiv*. https://doi.org/10.1101/2022.02.19.481103

Cole, M. W., Ito, T., Bassett, D. S., & Schultz, D. H. (2016). Activity flow over resting-state networks shapes cognitive task activations. *Nature Neuroscience*, *19*(12), 1718–1726.

Cole, M. W., Ito, T., Cocuzza, C., & Sanchez-Romero, R. (2021). The Functional Relevance of Task-State Functional Connectivity. *The Journal of Neuroscience: The Official Journal of the Society for Neuroscience*, *41*(12), 2684–2702.

Cole, M. W., Ito, T., Schultz, D., Mill, R., Chen, R., & Cocuzza, C. (2019). Task activations produce spurious but systematic inflation of task functional connectivity estimates. *NeuroImage*, *189*, 1–18.

Cole, M. W., Yarkoni, T., Repovs, G., Anticevic, A., & Braver, T. S. (2012). Global connectivity of prefrontal cortex predicts cognitive control and intelligence. *The Journal of Neuroscience: The Official Journal of the Society for Neuroscience*, *32*(26), 8988–8999.

Cybenko, G. (1989). Approximation by superpositions of a sigmoidal function. *Mathematics of Control, Signals, and Systems*, *2*(4), 303–314.

Damoiseaux, J. S., & Greicius, M. D. (2009). Greater than the sum of its parts: a review of studies combining structural connectivity and resting-state functional connectivity. *Brain Structure & Function*, *213*(6), 525–533.

de-Wit, L., Alexander, D., Ekroll, V., & Wagemans, J. (2016). Is neuroimaging measuring information in the brain? *Psychonomic Bulletin & Review*, *23*(5), 1415–1428.

Donahue, C. J., Sotiropoulos, S. N., Jbabdi, S., Hernandez-Fernandez, M., Behrens, T. E., Dyrby, T. B., Coalson, T., Kennedy, H., Knoblauch, K., Van Essen, D. C., & Glasser, M. F. (2016). Using diffusion tractography to predict cortical connection strength and distance: A quantitative comparison with tracers in the monkey. *The Journal of Neuroscience: The Official Journal of the Society for Neuroscience*, *36*(25), 6758–6770.

Esposito, R., Bortoletto, M., & Miniussi, C. (2020). Integrating TMS, EEG, and MRI as an approach for studying brain connectivity. *The Neuroscientist: A Review Journal Bringing Neurobiology, Neurology and Psychiatry*, *26*(5–6), 471–486.

Friston, K. J., Harrison, L., & Penny, W. (2003). Dynamic causal modelling. *NeuroImage*, *19*(4), 1273–1302.

Haak, K. V., Winawer, J., Harvey, B. M., Renken, R., Dumoulin, S. O., Wandell, B. A., & Cornelissen, F. W. (2013). Connective field modeling. *NeuroImage*, *66*(C), 376–384.

Hansen, K. B. (2020). The virtue of simplicity: On machine learning models in algorithmic trading. *Big Data & Society*, *7*(1), 205395172092655.
Hanson, C., & Hanson, S. J. (1996). Development of Schemata during Event Parsing: Neisser's Perceptual Cycle as a Recurrent Connectionist Network. *Journal of Cognitive Neuroscience*, *8*(2), 119–134.
Hearne, L. J., Mill, R. D., Keane, B. P., Repovš, G., Anticevic, A., & Cole, M. W. (2021). Activity flow underlying abnormalities in brain activations and cognition in schizophrenia. *Science Advances*, *7*(29), 2020.12.16.423109.
Henson, R. (2005). What can functional neuroimaging tell the experimental psychologist? *The Quarterly Journal of Experimental Psychology Section A*, *58*(2), 193–233.
Hornik, K., Stinchcombe, M., & White, H. (1989). Multilayer feedforward networks are universal approximators. *Neural Networks: The Official Journal of the International Neural Network Society*, *2*(5), 359–366.
Huth, A. G., Nishimoto, S., Vu, A. T., & Gallant, J. L. (2012). A continuous semantic space describes the representation of thousands of object and action categories across the human brain. *Neuron*, *76*(6), 1210–1224.
Hwang, K., Shine, J. M., Cole, M. W., & Sorenson, E. (2022). Thalamocortical contributions to cognitive task activity. *ELife*, *11*. https://doi.org/10.7554/eLife.81282
Ito, T. (2021). *Cognitive Information Transformation in Functional Brain Networks* (M. W. Cole, Ed.) [PhD]. Rutgers University-Newark.
Ito, T., Hearne, L., Mill, R., Cocuzza, C., & Cole, M. W. (2020). Discovering the Computational Relevance of Brain Network Organization. In *Trends in Cognitive Sciences* (Vol. 24, Issue 1, pp. 25–38). https://doi.org/10.1016/j.tics.2019.10.005
Ito, T., Kulkarni, K. R., Schultz, D. H., Mill, R. D., Chen, R. H., Solomyak, L. I., & Cole, M. W. (2017). Cognitive task information is transferred between brain regions via resting-state network topology. *Nature Communications*, *8*(1), 1027.
Ito, T., Yang, G. R., Laurent, P., Schultz, D. H., & Cole, M. W. (2022). Constructing neural network models from brain data reveals representational transformations linked to adaptive behavior. *Nature Communications*, *13*(673). https://doi.org/10.1038/s41467-022-28323-7
Kar, K., Ito, T., Cole, M. W., & Krekelberg, B. (2020). Transcranial alternating current stimulation attenuates BOLD adaptation and increases functional connectivity. *Journal of Neurophysiology*, *123*(1), 428–438.
Keane, B. P., Krekelberg, B., Mill, R. D., Silverstein, S. M., Thompson, J. L., Serody, M. R., Barch, D. M., & Cole, M. W. (2023). Dorsal attention network activity during perceptual organization is distinct in schizophrenia and predictive of cognitive disorganization. *The European Journal of Neuroscience*, *57*(3), 458–478.
Khan, A. G., & Hofer, S. B. (2018). Contextual signals in visual cortex. *Current Opinion in Neurobiology*, *52*, 131–138.
Krakauer, J. W., Ghazanfar, A. A., Gomez-Marin, A., MacIver, M. A., & Poeppel, D. (2017). Neuroscience needs behavior: Correcting a reductionist bias. *Neuron*, *93*(3), 480–490.
Kriegeskorte, N., Simmons, W. K., Bellgowan, P. S. F., & Baker, C. I. (2009). Circular analysis in systems neuroscience: the dangers of double dipping. *Nature Neuroscience*, *12*(5), 535–540.
Lee, J. H., Durand, R., Gradinaru, V., Zhang, F., Goshen, I., Kim, D.-S., Fenno, L. E., Ramakrishnan, C., & Deisseroth, K. (2010). Global and local fMRI signals driven by neurons defined optogenetically by type and wiring. *Nature*, 1–5.
Lee, J. H., Liu, Q., & Dadgar-Kiani, E. (2022). Solving brain circuit function and dysfunction with computational modeling and optogenetic fMRI. *Science*, *378*(6619), 493–499.
Li, L., & Spratling, M. (2023). Understanding and combating robust overfitting via input loss landscape analysis and regularization. *Pattern Recognition*, *136*(109229), 109229.

Lipworth, L., Friis, S., Mellemkjær, L., Signorello, L. B., Johnsen, S. P., Nielsen, G. L., McLaughlin, J. K., Blot, W. J., & Olsen, J. H. (2003). A population-based cohort study of mortality among adults prescribed paracetamol in Denmark. *Journal of Clinical Epidemiology*, *56*(8), 796–801.

Liu, Q., Ulloa, A., & Horwitz, B. (2022). The Spatiotemporal Neural Dynamics of Intersensory Attention Capture of Salient Stimuli: A Large-Scale Auditory-Visual Modeling Study. *Frontiers in Computational Neuroscience*, *16*, 876652.

Markov, N. T., Ercsey-Ravasz, M. M., Ribeiro Gomes, A. R., Lamy, C., Magrou, L., Vezoli, J., Misery, P., Falchier, A., Quilodran, R., Gariel, M. A., Sallet, J., Gamanut, R., Huissoud, C., Clavagnier, S., Giroud, P., Sappey-Marinier, D., Barone, P., Dehay, C., Toroczkai, Z., … Kennedy, H. (2014). A weighted and directed interareal connectivity matrix for macaque cerebral cortex. *Cerebral Cortex*, *24*(1), 17–36.

Mars, R. B., Passingham, R. E., & Jbabdi, S. (2018). Connectivity Fingerprints: From Areal Descriptions to Abstract Spaces. *Trends in Cognitive Sciences*, *22*(11), 1026–1037.

Martí-Juan, G., Sastre-Garriga, J., Martinez-Heras, E., Vidal-Jordana, A., Llufriu, S., Groppa, S., Gonzalez-Escamilla, G., Rocca, M. A., Filippi, M., Høgestøl, E. A., Harbo, H. F., Foster, M. A., Toosy, A. T., Schoonheim, M. M., Tewarie, P., Pontillo, G., Petracca, M., Rovira, À., Deco, G., & Pareto, D. (2023). Using The Virtual Brain to study the relationship between structural and functional connectivity in patients with multiple sclerosis: a multicenter study. *Cerebral Cortex (New York, N.Y.: 1991)*. https://doi.org/10.1093/cercor/bhad041

McClelland, J. L., & Rogers, T. T. (2003). The parallel distributed processing approach to semantic cognition. *Nature Reviews. Neuroscience*, *4*(4), 310–322.

McCormick, E. M., Arnemann, K. L., Ito, T., Hanson, S. J., & Cole, M. W. (2022). Latent functional connectivity underlying multiple brain states. *Network Neuroscience (Cambridge, Mass.)*, *6*(2), 570–590.

Mediano, P. A. M., Rosas, F. E., Luppi, A. I., Jensen, H. J., Seth, A. K., Barrett, A. B., Carhart-Harris, R. L., & Bor, D. (2022). Greater than the parts: a review of the information decomposition approach to causal emergence. *Philosophical Transactions. Series A, Mathematical, Physical, and Engineering Sciences*, *380*(2227), 20210246.

Mell, M. M., St-Yves, G., & Naselaris, T. (2021). Voxel-to-voxel predictive models reveal unexpected structure in unexplained variance. *NeuroImage*, *238*, 118266.

Mill, R. D., Bagic, A., Bostan, A., Schneider, W., & Cole, M. W. (2017). Empirical validation of directed functional connectivity. *NeuroImage*, *146*, 275–287.

Mill, R. D., Hamilton, J. L., Winfield, E. C., Lalta, N., Chen, R. H., & Cole, M. W. (2022). Network modeling of dynamic brain interactions predicts emergence of neural information that supports human cognitive behavior. *PLOS Biology*, 2021.01.26.428276.

Mill, R. D., Ito, T., & Cole, M. W. (2017). From connectome to cognition: The search for mechanism in human functional brain networks. *NeuroImage*, *160*, 124–139.

Mill, R., Flinker, A., & Cole, M. W. (2022). *Invasive human neural recording links resting-state connectivity to generation of task activity*.

Mueller, J. K., Grigsby, E. M., Prevosto, V., Petraglia, F. W., 3rd, Rao, H., Deng, Z.-D., Peterchev, A. V., Sommer, M. A., Egner, T., Platt, M. L., & Grill, W. M. (2014). Simultaneous transcranial magnetic stimulation and single-neuron recording in alert non-human primates. *Nature Neuroscience*, *17*(8), 1130–1136.

Naselaris, T., Kay, K. N., Nishimoto, S., & Gallant, J. L. (2011). Encoding and decoding in fMRI. *NeuroImage*, *56*(2), 400–410.

Osher, D. E., Saxe, R. R., Koldewyn, K., Gabrieli, J. D. E., Kanwisher, N., & Saygin, Z. M. (2016). Structural Connectivity Fingerprints Predict Cortical Selectivity for Multiple Visual Categories across Cortex. *Cerebral Cortex*, *26*(4), 1668–1683.

Palestro, J. J., Bahg, G., Sederberg, P. B., Lu, Z.-L., Steyvers, M., & Turner, B. M. (2018). A tutorial on joint models of neural and behavioral measures of cognition. *Journal of Mathematical Psychology*, *84*, 20–48.

Passingham, R. E., Stephan, K. E., & Kötter, R. (2002). The anatomical basis of functional localization in the cortex. *Nature Reviews. Neuroscience*, *3*(8), 606–616.

Pearl, J. (2009). Causal inference in statistics: An overview. *Statistics Surveys*, *3*(none), 96–146.

Pearl, J., & Mackenzie, D. (2018). *The Book of Why: The New Science of Cause and Effect*. Basic Books.

Reid, A. T., Headley, D. B., Mill, R. D., Sanchez-Romero, R., Uddin, L. Q., Marinazzo, D., Lurie, D. J., Valdés-Sosa, P. A., Hanson, S. J., Biswal, B. B., Calhoun, V., Poldrack, R. A., & Cole, M. W. (2019). Advancing functional connectivity research from association to causation. *Nature Neuroscience*. https://doi.org/10.1038/s41593-019-0510-4

Ritter, P., Schirner, M., McIntosh, A. R., & Jirsa, V. K. (2013). The Virtual Brain Integrates Computational Modeling and Multimodal Neuroimaging. *Brain Connectivity*, *3*(2), 121–145.

Rogers, T. T., & McClelland, J. L. (2014). Parallel Distributed Processing at 25: further explorations in the microstructure of cognition. *Cognitive Science*, *38*(6), 1024–1077.

Rozell, C. J., Johnson, D. H., Baraniuk, R. G., & Olshausen, B. A. (2008). Sparse coding via thresholding and local competition in neural circuits. *Neural Computation*, *20*(10), 2526–2563.

Rumelhart, D. E., Hinton, G. E., McClelland, J. L., & Others. (1986). A general framework for parallel distributed processing. *Parallel Distributed Processing: Explorations in the Microstructure of Cognition*, *1*(45–76), 26.

Rumelhart, D. E., Hinton, G. E., & Williams, R. J. (1986). Learning representations by back-propagating errors. In *Nature* (Vol. 323, Issue 6088, pp. 533–536). https://doi.org/10.1038/323533a0

Sacramento, J., Wichert, A., & van Rossum, M. C. W. (2015). Energy efficient sparse connectivity from imbalanced synaptic plasticity rules. *PLoS Computational Biology*, *11*(6), e1004265.

Sanchez-Romero, R., & Cole, M. W. (2021). Combining Multiple Functional Connectivity Methods to Improve Causal Inferences. *Journal of Cognitive Neuroscience*, *33*(2), 180–194.

Sanchez-Romero, R., Ito, T., Mill, R. D., Hanson, S. J., & Cole, M. W. (2023). Causally informed activity flow models provide mechanistic insight into network-generated cognitive activations. *Neuroimage*, *278*(120300), 120300.

Saxena, S., & Cunningham, J. P. (2019). Towards the neural population doctrine. *Current Opinion in Neurobiology*, *55*, 103–111.

Schultz, D. H., Ito, T., & Cole, M. W. (2022). Global connectivity fingerprints predict the domain generality of multiple-demand regions. *Cerebral Cortex*, *32*(20), 4464–4479.

Shen, X., Finn, E. S., Scheinost, D., Rosenberg, M. D., Chun, M. M., Papademetris, X., & Constable, R. T. (2017). Using connectome-based predictive modeling to predict individual behavior from brain connectivity. *Nature Protocols*, *12*(3), 506–518.

Sheth, S. A., Bijanki, K. R., Metzger, B., Allawala, A., Pirtle, V., Adkinson, J. A., Myers, J., Mathura, R. K., Oswalt, D., Tsolaki, E., Xiao, J., Noecker, A., Strutt, A. M., Cohn, J. F., McIntyre, C. C., Mathew, S. J., Borton, D., Goodman, W., & Pouratian, N. (2022). Deep brain stimulation for depression informed by intracranial recordings. *Biological Psychiatry*, *92*(3), 246–251.

Silver, D., Hubert, T., Schrittwieser, J., Antonoglou, I., Lai, M., Guez, A., Lanctot, M., Sifre, L., Kumaran, D., Graepel, T., Lillicrap, T., Simonyan, K., & Hassabis, D. (2018). A general reinforcement learning algorithm that masters chess, shogi, and Go through self-play. *Science (New York, N.Y.)*, *362*(6419), 1140–1144.

Smith, S. M., Fox, P. T., Miller, K. L., Glahn, D. C., Fox, P. M., Mackay, C. E., Filippini, N., Watkins, K. E., Toro, R., Laird, A. R., & Beckmann, C. F. (2009). Correspondence of the brain's functional architecture during activation and rest. *Proceedings of the National Academy of Sciences*, *106*(31), 13040–13045.

Smith, S. M., Nichols, T. E., Vidaurre, D., Winkler, A. M., Behrens, T. E. J., Glasser, M. F., Ugurbil, K., Barch, D. M., Van Essen, D. C., & Miller, K. L. (2015). A positive-negative mode of population covariation links brain connectivity, demographics and behavior. *Nature Neuroscience*, *18*(11), 1565–1567.

Tagamets, M. A., & Horwitz, B. (1998). Integrating electrophysiological and anatomical experimental data to create a large-scale model that simulates a delayed match-to-sample human brain imaging study. *Cerebral Cortex (New York, N.Y.: 1991)*, *8*(4), 310–320.

Tavor, I., Parker Jones, O., Mars, R. B., Smith, S. M., Behrens, T. E., & Jbabdi, S. (2016). Task-free MRI predicts individual differences in brain activity during task performance. *Science*, *352*(6282), 216–220.

Thompson, W. H., Nair, R., Oya, H., Esteban, O., Shine, J. M., Petkov, C. I., Poldrack, R. A., Howard, M., & Adolphs, R. (2020). A data resource from concurrent intracranial stimulation and functional MRI of the human brain. *Scientific Data*, *7*(1), 258.

Ulloa, A., & Horwitz, B. (2016). Embedding Task-Based Neural Models into a Connectome-Based Model of the Cerebral Cortex. *Frontiers in Neuroinformatics*, *10*, 32.

Ulloa, A., & Horwitz, B. (2018). Quantifying differences between passive and task-evoked intrinsic functional connectivity in a large-scale brain simulation. *Brain Connectivity*, *8*(10), 637–652.

Varoquaux, G., & Poldrack, R. A. (2018). Predictive models avoid excessive reductionism in cognitive neuroimaging. *Current Opinion in Neurobiology*, *55*, 1–6.

Vaswani, A., Shazeer, N., Parmar, N., Uszkoreit, J., Jones, L., Gomez, A. N., Kaiser, L., & Polosukhin, I. (2017). Attention is all you need. In *arXiv [cs.CL]*. arXiv. http://arxiv.org/abs/1706.03762

Wijeakumar, S., Ambrose, J. P., Spencer, J. P., & Curtu, R. (2017). Model-based functional neuroimaging using dynamic neural fields: An integrative cognitive neuroscience approach. *Journal of Mathematical Psychology*, *76*(Pt B), 212–235.

Yamins, D. L. K., & DiCarlo, J. J. (2016). Using goal-driven deep learning models to understand sensory cortex. *Nature Neuroscience*, *19*(3), 356–365.

Yamins, D. L. K., Hong, H., Cadieu, C. F., Solomon, E. A., Seibert, D., & DiCarlo, J. J. (2014). Performance-optimized hierarchical models predict neural responses in higher visual cortex. *Proceedings of the National Academy of Sciences of the United States of America*, *111*(23), 8619–8624.

Yan, T., Liu, T., Ai, J., Shi, Z., Zhang, J., Pei, G., & Wu, J. (2021). Task-induced activation transmitted by structural connectivity is associated with behavioral performance. *Brain Structure & Function*. https://doi.org/10.1007/s00429-021-02249-0